# Guiding, not Driving: Design and Evaluation of a Command-Based User Interface for Teleoperation of Autonomous Vehicles

Felix Tener

University of Haifa, Information System, felix.tener@gmail.com

Joel Lanir

University of Haifa, Information System, ylanir@is.haifa.ac.il

Autonomous vehicles (AVs) are rapidly evolving as an innovative mode of transportation. However, the consensus in both industry and academia is that AVs cannot independently resolve all traffic scenarios. Consequently, the need for remote human assistance becomes clear. To enable the widespread integration of AVs on public roadways, it is imperative to develop novel models for remote operation. One such model is tele-assistance, which promotes delegating low-level maneuvers to automation through high-level directives. Our study investigates the design and evaluation of a new command-based tele-assistance user interface for the teleoperation of AVs. First, by integrating various control paradigms and interaction concepts, we created a simulation-based, high-fidelity interactive prototype consisting of 175 screens. Next, we conducted a comprehensive usability study with 14 expert teleoperators to assess the acceptance and usability of the system. Finally, we formulated high-level insights and guidelines for designing command-based user interfaces for the remote operation of AVs.

## 1 INTRODUCTION

Recent advancements in technological fields such as computer vision, sensor fusion, and artificial intelligence have significantly accelerated the development of autonomous vehicles (AVs), positioning them as a transformative form of transportation [13]. Major automotive manufacturers and innovative startups are actively pursuing a wide range of advanced technologies to enable AVs to operate autonomously. However, the current capabilities of AVs reveal limitations in handling all possible road scenarios independently. Situations such as road construction, malfunctioning traffic signals, or congested intersections may hinder an AV's autonomous operation [9,14,41,64]. As a result, there is a prevailing consensus among both academic and industry experts that, for the foreseeable future, AVs will encounter complex traffic scenarios that necessitate remote human intervention [4,8,22,56].

A promising approach to effectively address these challenges and facilitate the widespread deployment of fully autonomous vehicles is teleoperation. AV teleoperation involves a remote human operator (RO) who oversees and manages the vehicle's actions from a distance. In situations where a vehicle encounters difficulties, the RO can be called upon to assess the situation and guide the vehicle until the issue is resolved. The RO, typically stationed in a remote operation center, may oversee multiple AVs during a single teleoperation shift. Several teleoperation systems for AVs are currently in operation and continue to be refined by various automotive companies [27,45]. These systems predominantly use Tele-driving, in which the RO continuously controls the vehicle using a steering wheel and pedals (Figure 1, left). However, manual remote control of a vehicle has proven to be exceedingly challenging [63]. One significant difficulty arises from the physical separation between the RO and the vehicle, which prevents the operator from sensing the forces exerted on the vehicle or hearing the surrounding sounds. Additionally, latency is a concern, as a large volume of data must be transmitted from the AV to the RO over the network [21,68].

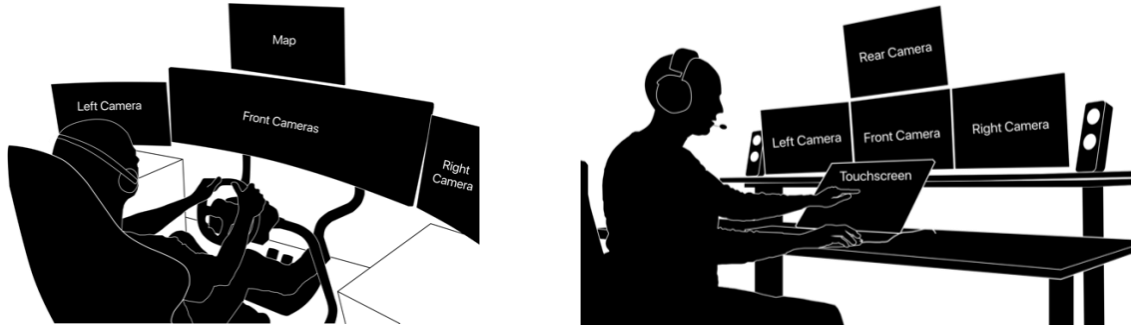

Figure 1: Left – schematic drawing of teleoperation via a steering wheel and pedals (tele-driving). Right –teleoperation via high-level commands (tele-assistance).

*Tele-assistance* introduces a different paradigm in which humans provide guidance-level input to an automated system. In this model, the RO delegates the execution of low-level maneuvers to the AV, issuing high-level directives through a specialized interface [19]. The operator still receives and views the video feed and information from the remote vehicle; however, control is exerted through interface commands rather than direct driving (Figure 1, right).

Because in tele-assistance, the RO does not have to drive the AV continuously, this novel paradigm offers many advantages over manual vehicle teleoperation. These advantages include shorter teleoperation sessions, reduced RO's cognitive load, overcoming of the latency barrier [53,63], easier managing of diverse vehicle types, and increased safety [65]. Previous studies have investigated in which road scenarios AVs might require remote human assistance [3,5,46,48,64]. Following these studies, Tener and Lanir defined a set of discrete high-level commands that can be used by ROs to make critical decisions while delegating detailed maneuvers to automated systems [65]. Building upon these foundations and other works [9,16,36,41,45], this research aims to answer the fundamental question: "How do we design a command-based tele-assistance interface for the teleoperation of AVs?". Our work synthesizes the accumulated knowledge and utilizes the *Research through Design* [69] approach to develop a first-of-its-kind teleassistance user interface. Specifically, we designed a command-based interface where the remote operator provides high-level guidance to autonomous vehicles via a touch-based surface. While the proposed design paradigm is generic and can be relevant to a diverse range of AVs, we focused on AVs in urban areas, such as robotaxis and automated shuttle buses.

To assess this interface, we simulated three representative edge-case scenarios and incorporated these, along with the identified high-level commands and other interaction techniques [17,35,60], into a high-fidelity interactive prototype comprising 175 distinct screens. We performed a comprehensive usability study on the prototype with 14 experts, all of whom possess substantial experience in the remote operation of ground vehicles. The primary contributions of this paper are as follows:

1. The development of a novel command-based tele-assistance user interface for a touch surface.
2. An evaluation of the tele-assistance interface with 14 experienced teleoperators.
3. A set of insights for the future design of tele-assistance command-based user interfaces.

## 2 RELATED WORK

### 2.1 Autonomous vehicles and remote driving

Autonomous vehicle research, though still in its early stages, is advancing rapidly due to significant innovations in perception and decision-making technologies [2,55]. Consequently, numerous studies have investigated the dynamics between humans and AVs. Some of these studies have concentrated on AV-pedestrian interactions [47,57], looking into the understanding, perceptions, and trust associated with AVs. Other research has focused on the interactions between AVs and their passengers [10], with particular emphasis on shared autonomous vehicles [51]. Additionally, there has been considerable attention on the development and evaluation of external human-machine interfaces (eHMIs) for AVs [7,8,11,28], highlighting their importance in facilitating effective communication between humans and AVs.

Along with the above-mentioned explorations, researchers acknowledged that despite the advancements in artificial intelligence and machine learning, AVs will inevitably encounter limitations when navigating edge-case road scenarios. Multiple research works have analyzed and classified disengagement scenarios [9,14,16,41,45] and take-over requests (TORs)[48]. Disengagements mostly happen in Level 2 ("Partial Driving Automation") or Level 3 ("Conditional Driving Automation") of automation, as defined by the SAE[1] [58], when the vehicle returns to manual control, or the in-vehicle driver feels the need to take back the wheel from the AV decision system. Expanding the above-mentioned research to higher levels of automation (i.e., when there is no driver in the vehicle), intervention use cases might happen when AV's sensors and algorithms fall short [64]. For example, an AV might struggle to accurately identify objects (e.g., a shade vs. a real object), confront unfamiliar circumstances (e.g., road construction), or face perception difficulties due to environmental factors (e.g., heavy rain).

While humans might also encounter challenges in such edge cases, they are often adept at interpreting complex or novel situations. Consequently, a human remote operator can manage situations that pose significant challenges for an AV. Additionally, certain aspects of vehicle operation may necessitate human intervention to ensure regulatory compliance (e.g., crossing a continuous separation lane to bypass an animal blocking the road). As such, the presence of an RO is crucial for effectively addressing these edge-case scenarios.

The consensus in the industry and academia is that autonomous fleets will be overseen and managed from remote operation centers [15,34]. Such centers might accommodate multiple ROs. Several research works investigated such teleoperation centers. For example, Feiler et al. [15] conceptualized a control center for an AV fleet and proposed several service categories. Kettwich et al. [34] defined the requirements of future control centers in public transport, and Franke et al. [20] designed an interface for a control center of autonomous buses.

Remote driving, or Tele-driving, is the most common way to operate vehicles remotely today. It has been widely considered and discussed commercially and academically [1,44,53]. This method involves using components such as a steering wheel, pedals, and multiple screens displaying real-time video feeds from various onboard cameras (front, rear, right, and left perspectives). These elements enable ROs to assume complete remote control of an AV, managing its low-level maneuvers. However, tele-driving is a very challenging task. These challenges have been extensively examined in the literature [23,30,63,68]. A primary source of difficulty in remote vehicle operation arises from the physical disconnection between the RO and the AV, compounded by network latency. This physical separation renders it exceedingly challenging for the operator to accurately gauge acceleration, speed, road gradients, and tactile feedback from the vehicle's pedals. Moreover, the absence of reliable video transmission can render tele-driving infeasible. Additional challenges in tele-driving stem from human cognitive and perceptual limitations, such as increased cognitive load, difficulty in correctly estimating acceleration, speed, depth perception, and the absence of auditory cues.

---

[1] SAE – Society of Autonomous Engineers (https://www.sae.org/).

Recent research endeavors have begun to examine the prerequisites for interfaces centered around teleoperation and delineate the design space for AV teleoperation interfaces [26]. The work of Graff and Hussmann [25,26] has been pivotal in compiling an extensive set of user requirements for AV teleoperation, which includes detailed information on the vehicle, its sensory systems, and the surrounding environment (e.g., vehicle positioning, operational status, weather conditions, etc.). Nonetheless, it remains evident that latency will persist as a critical bottleneck in the near and foreseeable future [53], and bridging the physical disconnect between operator and vehicle may necessitate substantial engineering innovations and financial investments [5]. Consequently, the evolution of novel teleoperation paradigms is imperative.

**2.2 Tele-assistance**

Several research works have defined two primary teleoperation modalities with which ROs can assist AVs from within teleoperation centers: *Direct Control* and *Indirect Control* [4,43]. In direct control (tele-driving), the RO continuously controls the AV using a steering wheel and pedals. In indirect control (tele-assistance), the RO does not take full control of the AV; instead, the RO delegates low-level maneuvers to the AV without the need to directly manage the vehicle's movements.

The implementation of tele-assistance presents multiple advantages over manual vehicle operation. Firstly, the use of remote directives markedly shortens teleoperation sessions, as issuing succinct commands such as "bypass from left" or "wait" is considerably more expedient than manually steering the vehicle [65]. Secondly, by decoupling the RO from the AV's low-level systems, such as the steering wheel and pedals, ROs gain the capacity to manage diverse vehicles - differing in size, width, and type - and various fleets, including private cars, shuttles, and trucks, through standardized commands. This approach streamlines the adaptation process and eases the learning curve, as ROs are not required to develop new mental models when transitioning between different teleoperated vehicles [63]. Thirdly, tele-assistance offers a critical enhancement in safety. Data from the U.S. Department of Transportation reveals that human error is responsible for 94 percent of accidents in the United States [30]. Recent findings from Waymo further highlight that the autonomous vehicle crash rate - 0.41 incidents per million miles of driving - is significantly lower than the human driver crash rate of 2.78 per million miles [37]. Thus, delegating low-level maneuvers to autonomous systems holds substantial promise for improving AV safety. Lastly, tele-assistance has the potential to reduce the cognitive load imposed on ROs by tele-driving interfaces, which typically demand a high level of attention [63]

Several studies proposed various components of tele-assistance. For example, Mutzenich et al. [50] argue that the RO user interface should facilitate tasks such as communication with service providers, passengers, and fleet management centers. Fennel et al. [17] propose a method to delineate an AV's intended path using an expansive haptic interface on which a user physically walks. Similarly, Schitz et al. [60] address this challenge by defining a collision-free corridor within which the AV can compute a safe trajectory. Kettwich and Schrank et al. [35,61] developed and evaluated a tele-assistance user interface specifically designed for the teleoperation of automated public transport shuttles that follow pre-determined routes. This interface allows the teleoperator to select a particular shuttle, provide assistance by interacting with third parties or passengers, and adjust the shuttle's trajectory through a touch screen. Flemisch [19] and Kauer [33] introduce an approach where the RO communicates with the AV via maneuver-based commands, which are then executed by the vehicle's automated systems. Aramrattana et al. [18] suggested a roadmap towards remote assistance, where they suggest exploring various tele-assistance aspects (e.g., standards and regulations). Finally, Tener and Lanir [65] devised a comprehensive high-level command language for the teleoperation of AVs, which is composed of discrete commands, such as "Bypass from Right" or "Slide."[10]

While these studies have explored diverse indirect control aspects, to our knowledge, no research work has designed a comprehensive command-based tele-assistance interface for the remote operation of AVs. The current research builds upon the above-mentioned research works, specifically [19,43,65], designing and evaluating a high-fidelity interactive prototype with which an RO can guide an AV using mainly indirect control techniques applied to a touch surface.

## 3 THE TELEOPERATION CONTEXT AND DESIGN CONCEPT

To support AVs in edge-case situations, several research groups have explored the concept of a control center for automated vehicle fleets [15,34], and various companies (e.g., DriveU, Phantom Auto) build teleoperation stations to populate such centers. When an AV encounters a road scenario that it cannot handle autonomously, it requests remote operation assistance from the teleoperation center. Such a request is automatically assessed and evaluated according to its various characteristics (urgency, vehicle type, etc.) and enters a queue of requests that should be resolved. Similarly to teleoperation requests, each RO within the center will have various characteristics. For instance, some ROs might be more qualified to operate autonomous buses than others. Another example is RO's vigilance levels, which can be tracked and assessed using physiological measurements in real-time [49]. Once the next available and suitable RO is ready for the next teleoperation session, he or she will be automatically matched with the next-in-line teleoperation request and receive a notification in their station.

After receiving the notification, the RO will assess the remote environment and choose the most appropriate method to resolve the situation, depending on the use case. For instance, if the AV is required to merge into a busy intersection, the RO might take full remote control of the vehicle and perform the needed maneuver (tele-driving). Alternatively, if there is a static obstacle on the road, the RO could simply provide high-level guidance to the AV (tele-assistance). If the remote scenario cannot be addressed from afar (e.g., flat tire), the RO should be able to call a field team, contact the passengers, inform them about the situation, and book an alternative shuttle [50,64].

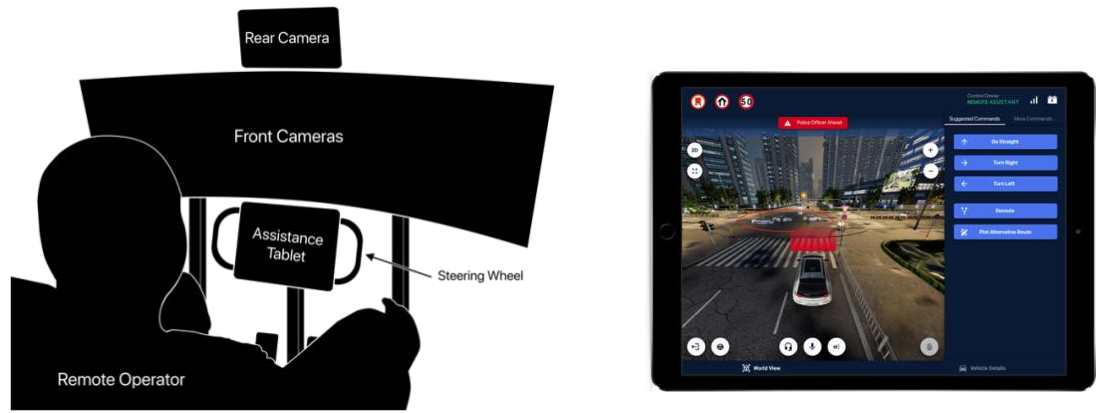

Figure 2: Left - A proposed teleoperation station that incorporates both tele-driving (a steering wheel and pedals) and tele-assistance (touch-based tablet) components, as well as screens with real-time video feeds from front and back cameras. Right - A tablet based tele-assistance user interface that incorporates contextual high-level commands.

Resolving possible edge cases using tele-assistance is likely faster and safer than remote driving [30,65]. At the same time, it is also important to enable operational flexibility and allow ROs to drive AVs remotely if necessary. Thus, we propose enabling both operational modes (tele-driving and tele-assistance) in the same teleoperation station. Therefore, our teleoperation station design includes a steering wheel, pedals, and a tablet-based assistance interface (Figure 2 - left). Additionally, the station should have several screens showing a real-time video feed from the back, front, and side vehicle-mounted cameras. Finally, data from other sensors, such as LIDAR, GPS, radar, sonar, etc., should be transmitted and intuitively presented to the RO.

An assumption we made in our design is that when an AV encounters a situation that it cannot resolve autonomously, it will, in most cases, still recognize the reason for the disengagement (e.g., an obstacle blocking the road) [42,55]. Based on the work by Tener and Lanir [65], each recognized disengagement reason can be mapped to a set of high-level commands, using which the edge case can be addressed. Such commands will be context-dependent and dynamically change based on the use case. For example, the RO can use the "Bypass from Left" command when the AV encounters a tree branch that blocks its route. In the current work, we designed a command-based dialog, which unfolds between the RO and the AV after selecting the desired command. In our design, we incorporated examples of such commands in the tele-assistance interface (Figure 3, Area 3).

A tele-assistance interface should support not only high-level commands [65] but also employ various tools to draw a desired driving path, make a selection, and add additional data to an object within a remote scene [43]. To support all the above affordances, we suggest using a tablet-based user interface (Figure 2 - right). Touch interfaces are not only ubiquitous, intuitive, and quickly adopted as a command tool but also were used by other researchers [35,37] in their design of tele-assistance stations.

## 4 USER INTERFACE DESIGN

This section describes the main design considerations, decisions, and flows made in designing the tele-assistance user interface. A video exemplifying the user interface through several use cases can be seen in the supplementary material.

### 4.1 Prototype

We selected three representative edge-case scenarios based on the categorization suggested by Tener and Lanir [65]: (1) Police blocking the road, (2) Integration into heavy traffic, and (3) Static obstacles blocking the road. We also adopted the commands that were found to be effective for the resolution of these edge cases in the same study. We used a specialized simulation platform[2], to create 30-60 seconds-long simulations depicting the upper-mentioned use cases. We then used the simulations to take multiple screenshots to animate the AV's behavior based on a selection of each command. Finally, we integrated the simulated screens and the elicited commands into a high-fidelity interactive prototype created using Sketch[3] and InVision[4] prototyping software. The final prototype comprises 175 screens.

### 4.2 Main screen topology

Figure 3 presents the main screen of the tablet-based tele-assistance UI. It is divided into four major areas: the upper status bar, the remote environment representation, the contextual command menu, and the bottom navigation bar.

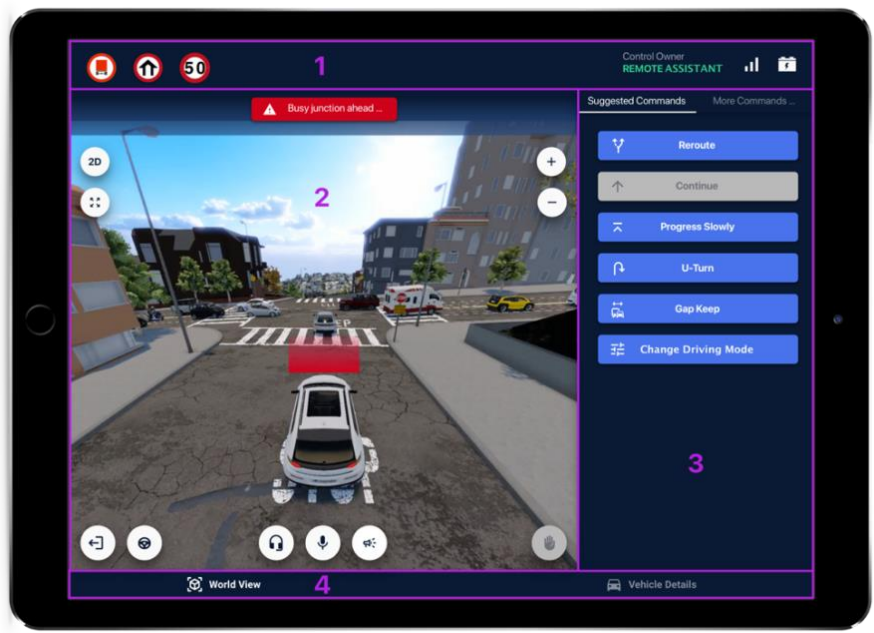

Figure 3: Tele-assistance UI – main screen topology.

#### 4.2.1 Upper-status bar

The status bar (Figure 3, Area 1) is located in the upper part of the main screen. Its primary purpose is to notify the RO about (1) The current or future road situation, (2) The current state of the AV, and (3) The current control owner. The status bar is roughly divided into two major sub-areas.

On its left part, the user can see the status of the current road situation. For example, she can see which road rules apply in a specific road section (e.g., speed limit) and various dynamic road notifications such as road construction zones, etc.

---

[2] https://www.cognata.com/
[3] https://www.sketch.com/
[4] https://www.invisionapp.com/

Such information is especially important because the RO usually enters a session without context and may not know the history of events that led the AV to its current situation [52]. Thus, road signs from previous road sections or notifications about future events (e.g., traffic jam ahead) provide the RO with the needed context.

The right part of the status bar encompasses essential system notifications (e.g., AV's battery life) as well as the current control owner of the AV. There are three major control states: (1) Vehicle, (2) Remote Assistant, and (3) Remote Driver (Figure 4). When the AV progresses autonomously, the control state is "Vehicle", when the AV is continuously driven via a steering wheel and pedals, the control state is "Remote Driving," and when the AV is operated using discrete commands, the status is "Remote Assistant". Since humans and AI alternately control AV movements in tele-assistance, it is essential to reflect control ownership of the RO at all times.

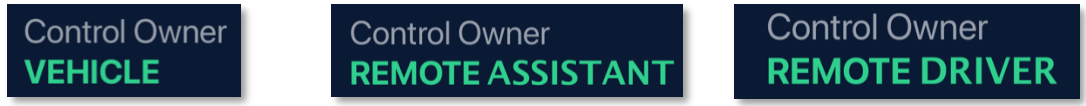

Figure 4: AV control owners (elements from the tele-assistance UI).

### 4.2.2 Remote environment representation

This screen area (Figure 3, Area 2) represents AV's remote environment in a third-person perspective. Sensor fusion techniques might achieve such a world-view representation by integrating data from various AV sensors (LIDAR, radar, cameras, GPS, etc.) [62,67].

We also propose to place several controls on top of the video feed. The buttons on the top right and top left control the zoom and the point of view of the remote scene, while the buttons at the bottom of the screen are commands that are frequently used by the RO and are not scene-contextual. Table 1 describes each command button.

Table 1: Frequently used UI commands.

| # | Command | Command Description |
|---|---|---|
| 1 | 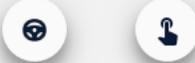 | **Transition to driving mode / assistance mode** - according to our conceptual model, the AV can be controlled autonomously, via tele-driving, or via tele-assistance. The images on the left show a toggle button that transitions the user from tele-assistance to tele-driving (left) and vice versa (right). |
| 2 | 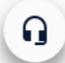 | **Contact passenger / 3rd party / external human** - One of the most important affordances of a tele-assistance UI is communication with 3rd parties (e.g., a fleet management center[5] [50,65], other road users (e.g., a compound guard), and passengers. This command button enables the initiation of such a communication. |

---

[5] **Fleet management center (FMC)** – a center (physical location) that is used to monitor and assist various vehicles of the same client (e.g., bus company). Stakeholders within the FMC are responsible for a smooth and positive end-to-end experience of the company's clients (e.g., Uber riders).

| 3 | 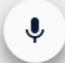 | **Microphone** – This command enables activating an external microphone (and a listening system), which affords the RO to vocally communicate with somebody who is located in a close proximity to the AV (e.g., a police officer). |
| 4 | 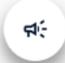 | **Honk** - pressing on this button will remotely activate the honk. |
| 5 | 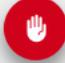 | **Stop** – activating this command will stop the AV. |
| 6 | 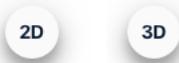 | **Camera perspective change** - the RO can change camera perspectives from 2D to 3D and vice versa using this button command. The 2D perspective is better for path planning while the 3D perspective preferable for high-level guidance. |
| 7 | 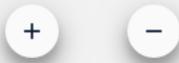 | **Zoom in / out** - these buttons enable incremental zoom (in and out) of the remote scene. |
| 8 | 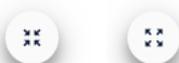 | **Changing scale of the remote scene** - in some scenarios (e.g., rerouting) it is necessary to zoom out significantly to view the whole route of the AV (from the beginning to the end). In such cases zooming in incrementally might be tedious for the user. Therefore, we enabled changing the scale of the remote scene at once. |
| 9 | 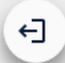 | **Log out** - this button logs the user out of the system. |

### 4.2.3 Contextual commands menu

The command menu (Figure 3, area 3) has two major sub-areas (tabs): (1) Contextual commands and (2) All commands. The "Contextual Commands" area includes high-level commands associated with the current detected disengagement scenario [64,65]. The underlying assumption is that these commands are presented to the RO by an assisting AI agent, which can decipher the remote scenario and then associate it with relevant high-level commands that can address it. A similar approach was demonstrated by Trabelsi et al. [66]. Additionally, this area includes dialog-specific actions necessary when providing an end-to-end resolution of a use case. For instance, the commands "Find Alternative Route," "Cancel," and "Confirm" will appear on the screen (Figure 5) after the "Reroute" command is selected in the previous screen.

The "All Commands" area allows users to access all commands anytime. It includes all the existing interface commands, divided into logical clusters to ease their finding. Possible grouping of the commands can be: (1) Frequently used, (2) Route control (e.g., "Reroute"), (3) Movement pace control (e.g., "Progress Slowly"), (4) Other autonomy modes (e.g., "Apply an emergency routine"), (5) View improvement (e.g., "Zoom"), (6) Withing-between lane placement (e.g.,

"Snap to the Right"), and other categories, [65]. Another important feature that can enable quick finding of each command is a search button to allow searching through all commands across categories. This area is especially useful in cases where the AI fails to diagnose the reason for the intervention, and the RO is required to comprehend why she was called to assist the AV and use her learned knowledge regarding the possible resolution options.

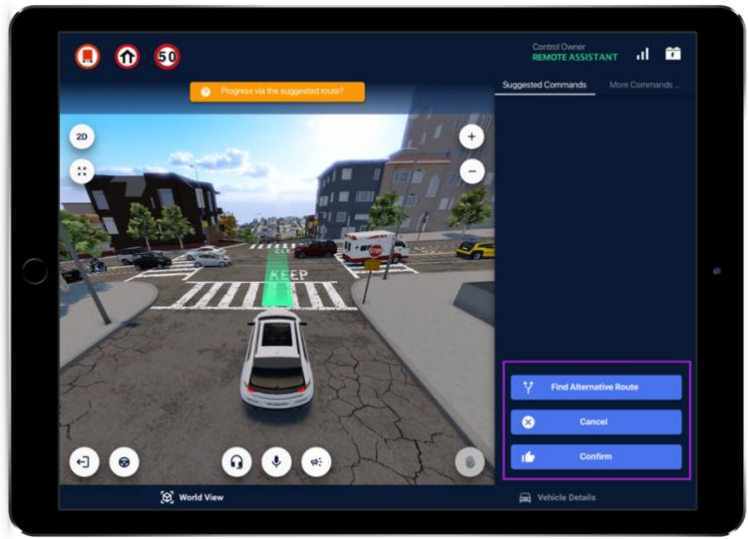

Figure 5: Tele-assistance UI - dialog-dependent commands. The commands "Find Alternative Route," "Cancel," and "Confirm" will appear on the screen after the "Reroute" command is selected in the previous screen (Figure 3).

### *4.2.4 Bottom navigation bar*

The bottom navigation bar (Figure 3, Area 4) is used for navigation within the tele-assistance application. Currently, the navigation bar includes two tabs. The first tab presents all the above areas and is expected to be the virtual space where users spend most of their time. The second tab encapsulates essential, yet secondary, details such as the vehicle's details (e.g., vehicle ID), information about the weather in the AV's surroundings, the location of the AV in space (pitch, yaw, roll), etc. Additional tabs may be added to the UI in future design versions.

### 4.3 Interactions

Several interaction types were explored in the current design:

1. **Button clicks** - The primary paradigm that was explored is the use of discrete high-level commands (e.g., "Bypass from Left") [65], via which the RO can make high-level decisions and delegate low-level maneuvers to the automation [19,31]. In our prototype, these commands are activated by the RO through button clicks.
2. **Path plotting** - Several research works have explored various ways to plot a desired trajectory for an AV [17,35,43,60]. The current design allows ROs to draw a route by placing tracing points on a 2D map (Figure 6, left).
3. **Interaction with AI** - humans and AI alternately control AV movements in tele-assistance. One example of AI assisting humans is suggesting new routes when the AV encounters an obstacle. In such a case, the RO may progress by selecting one of the proposed routes [43].

4. **Selection of objects inside the remote environment** - the user should be able to select objects (e.g., obstacles, other vehicles) in the remote scene. One possible example of such an interaction is the "perception modification concept" presented by Majstorovic et al [43]. In this interaction, the human operator selects an object within the remote scene and identifies whether the selected object is a "free space," "static object," or "dynamic object." (Figure 7, right).
5. **Adding human-sensor data to the remote scene** - in some cases, humans can comprehend the remote situation better than the AV. For instance, the AV might observe a dynamic obstacle ahead of the vehicle but not distinguish whether it is a falling rock or a flying plastic bag. In such cases, the RO should be able to select an object within the remote scene (see Figure 7) and add additional data to it [43].

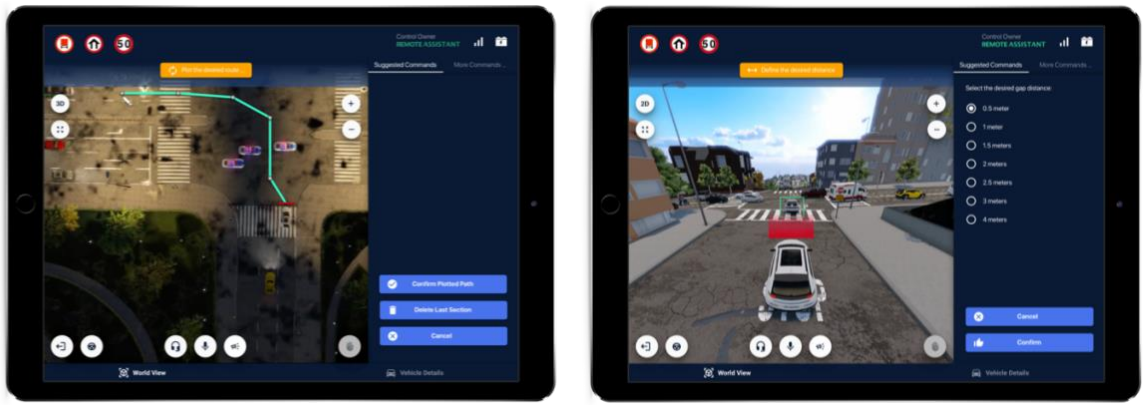

Figure 6: Left - plotting an alternative route. Right – selecting an object in the remote scene.

### 4.4 On-screen augmented reality (AR) overlays and notifications

On-screen overlays, added on top of the photo-realistic representation of the remote environment (Figures 3,5,6, and 7), play a crucial role in the design of the tele-assistance UI because they provide real-time contextual feedback to the RO [63]. In addition, being located in the center of the screen, these augmentations provide the RO with essential information without requiring her to move her gaze to the screen's periphery. In our prototype, in addition to the object selection and path plotting that were previously described, we designed three major types of overlays:

1. **Obstacle marking** – Marking the reason for intervention increases the RO's situation awareness and reduces the RO's reaction time [63]. In the right image of Figure 2 we demonstrate one possible implementation of the above concept - a red circle around police vehicles that block the road.
2. **Brakes pressing visualization** - to increase RO's feeling of control, we chose to visualize the brakes pressing of the AV using two methods: (1) Showing a virtual red wall in front of the AV and (2) Activating the brake lights on the representation model of the vehicle (Figures 3). [71]
3. **Projection of AV's future trajectory** - To visualize the AV's trajectory and distance to the stoppage, we used a virtual path that starts in front of the operated vehicle and continues forward depending on the AV's current speed and future trajectory. The faster the AV moves, the longer such a projection becomes (Figure 5). A similar approach was adopted by Waymo when designing its passenger's application [24].

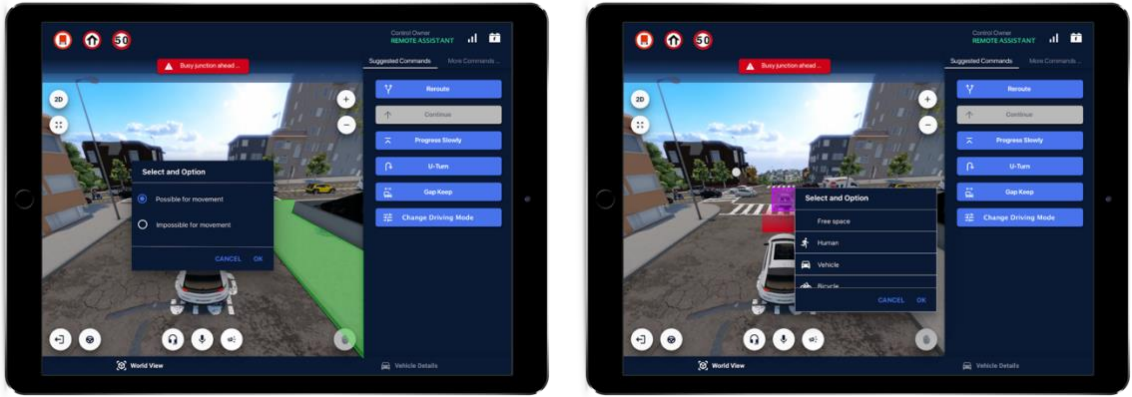

Figure 7: Object selection and adding human sensor data. Left – static object repurposing (allowing the AV to drive on the pavement). Right – dynamic object identification.

### 4.5 Notifications

Notifications are another way in which the AV can provide various types of information (warnings, guidance, etc.) to the RO, and we envision the notifications to be sent by an assisting agent, similar to the TeleOperator Advisor proposed by Trabelsi et.al [66]. In our design suggestion, there are three major types of notifications: (1) Alerts, (2) Progress notifications, and (3) Success notifications. Alerts appear in red and generally indicate a problem that AV's autonomous systems cannot resolve (Figure 3, upper part in area 2). Progress notifications appear in yellow-orange (Figure 5) and can be divided into two sub-categories: (a) Status indication, which informs the RO about the AV's current status (e.g., "Progressing"), and (b) Guidance, which guides the user to take a certain action to progress to the next step (e.g., "Select the desired mode"). Finally, we used success notifications, which symbolize the end of the interaction and appear in the interface in green color and an appropriate ("V") icon.

## 5 EVALUATION

To evaluate the user interface and gain insights into the design of tele-assistance interfaces, we conducted a usability study with 14 participants with extensive experience in ground vehicle remote operation. In our research, in addition to measuring the system's usability [29,39] and quantifying its acceptance [38], we aimed to gain a high-level understanding of using discrete commands as a teleoperation method for edge case scenarios. The study was performed remotely via the Zoom conferencing tool and was executed under ethical approval from our institution's ethical review board.

### 5.1 Methodology

#### 5.1.1 Participants

Fourteen participants (9 identifying themselves as female and five as male) aged 23 to 45 (M = 32.46, SD = 7.02) participated in the study. Participants were initially recruited among experts of a large consortium focusing on developing command-and-control systems in autonomous vehicles in urban areas. Additional participants were recruited using snowball-like sampling. Eleven participants had bachelor's degrees or higher. Twelve interviewees had extensive teleoperation experience with various types of unmanned ground vehicles (UGVs), and two participants were involved in developing sophisticated teleoperation systems while closely collaborating with teleoperators. All participants agreed to participate in the study, signed an informed consent form, and received a digital voucher as symbolic compensation for their contribution. Specific details on the participants' teleoperation experience are provided in Table 2.

Table 2: Participants' AV-related experience.

| Participant | Current / Most Relevant Role | Teleoperation Experience Description |
|---|---|---|
| RO1 | UGV teleoperator | Acted as an army UGV (Tomcar[6]) teleoperator for 2 years. The UGV could be operated using a steering wheel and pedals, as well as using discrete commands (stop, play, and define a route) |
| RO2 | A senior test manager and an autonomous driver in a company deploying autonomous shuttles | Planned and executed various tests for automotive and smart transportation technologies. Has 3+ years of teleoperation experience |
| RO3 | A QA tester in a company which develops unmanned ground systems | Operated a UGV (Tomcar and Ford-350) during her army service and was also employed as a QA tester in a company that developed unmanned ground systems |
| RO4 | Army officer | Operated a Tomcar-based UGV during her army service using a steering wheel and pedals. Also used a keyboard to activate lights and create sounds |
| RO5 | Army UGV teleoperator, officer and a training professional | Teleoperated a UGV during an army service of 3 years and trained other ROs. The AV could be driven semi-autonomously (follow a predefined path) and manually (via a steering wheel and pedals) |
| RO6 | Program lead in a global technology and engineering company | In the last 3 years worked on the product and development aspects of teleoperation shuttles and busses (first and last mile mobility). Overall employed in the autonomous field for around 6 years |
| RO7 | AVM Technician at a large software corporation | A Tomcar-based unmanned ground border patrol vehicle commander in the infantry. Used a steering wheel and pedals to control the vehicle |
| RO8 | Product marketing manager at a digital medical technology company | RO and commander of a Tomcar during her army service. The UGV was operated using a steering wheel and pedals from Logitec. RO8 controlled the UGV's cameras using a joystick |
| RO9 | Test engineering team leader in a company that provides robots as a service | Underwent a teleoperation course and has several dozens of teleoperation hours of an autonomous shuttle controlled via an Xbox joystick. Employed as a test engineer for 3 years |
| RO10 | Medical student | Completed a course of UGV operation and a Jeep driving course. Operated a UGV using a steering wheel and pedals for 1.5 years. |
| RO11 | Senior engineer in a large corporation that develops autonomous service solutions | In the last five years works on integration of a fleet management system with various autonomous systems that require teleoperation: AVs, buses, shuttles, robots, etc. |
| RO12 | Graphic designer at a design agency | Operated a UGV (Ford-350) during her army service (2 years) using steering wheels and pedals |
| RO13 | Student | Commanded over operation of various UGVs in during her army service. In particular, a Ford-350, dedicated for border protection and information gathering |
| RO14 | Founder of an autonomous vehicle and robotic system engineering consulting company | Founded an AVs and robotics branch, where he spent researching, developing, and deploying AVs for real world applications for 10 years. Later, co-founded a company for teleoperation of AVs |

---

[6] https://tomcar.com/

*5.1.2 Procedure*

The study was conducted remotely via the Zoom video-conferencing tool. The participants were given access to the interactive prototype via an Internet link, shared their screens during the usability test, and were encouraged to follow a think-aloud protocol [54] while performing the test tasks. Participants saw iPad-based screens (Figure 3) in their browsers and could click through them using predefined hot spots developed in InVision [32]. All sessions were recorded and lasted 90 minutes on average. After obtaining an informed consent form and a demographic questionnaire, we gave our users a brief introduction describing the purpose of the study. Then, we requested the users to perform a screen exploration. During this stage, participants were encouraged to walk through the prototype freely while clicking buttons and explaining what they saw in the interface to the researchers. After validating that the users correctly understood the remote situation, they were asked explicitly about the discrete commands, the augmented reality overlays, the traffic signs, the indicators, and the menu tabs (Figure 3). Next, users were tasked with resolving each of the three road scenarios (see section 5.1.1) using different commands. For instance, if users selected the "Bypass from Left" command to overtake a tree, they were also requested to resolve the same edge case using other commands. After completion of the tasks, users were asked to fill out a post-study system usability questionnaire (PSSUQ [39]) and a human-machine interface acceptance questionnaire (VDL[38]). At the end of the study, we conducted a summative interview.

*5.1.3 Data Analysis*

First, we manually transcribed the participants' responses from the recorded videos. Next, to analyze the collected qualitative data, we employed elements of Braun and Clark's thematic analysis approach [6]. For each scenario, we created a mapping between the discussed UI component (e.g., the "Control Owner" indicator) and the participants' actions and comments about this particular component (e.g., P13 thought that the transition of control was to the passenger inside the AV). We combined similar responses, eliminating redundancies, counted how often each comment appeared, and indicated which participant mentioned it. A similar method was applied to the post-study interview questions.

## 5.2 Results

We first present the main themes that emerged from the evaluation, focusing on insights regarding the use of high-level commands and key considerations for designing such an interface. Usability issues, which are more specific to the interface design, are detailed in the appendix. Finally, we present the quantitative evaluations derived from the questionnaires.

*5.2.1 Key insights*

Most participants thought the UI was useful and efficient. They did not think there were unnecessary features in the interface and particularly liked the screen topology, the suggested commands, the chosen camera perspectives, the visibility of contextual traffic signs, and the white rounded controls placed on top of the video feed. RO2 said he *"...liked the overall structure of the screen, the location of the buttons, and the separation between 'Suggested' and 'More Commands'..."*. RO9 shared that he *"... liked the transition between 2D and 3D. It's really helpful ..."*. Finally, RO14 emphasized that *she "... really liked the traffic signs because you are landing [to the UI] from nowhere [and have no context without the traffic signs]... This is super useful even in the Advanced Driver Assistance systems (ADAS), especially abroad ..."*. Saying the above, multiple usability issues were raised, and high-level insights were gained during the study. In the following subsections, we mainly focus on insights that can be generalized to future designs of tele-assistance user interfaces.

*Guiding an AV Using Discrete Commands*

All but one participant thought that controlling an AV using discrete high-level commands was a realistic and desirable approach. RO2 said, *"... yes because (in the use cases I've seen) it prevents human error... it seems reasonable and desirable that the AV continue to move autonomously based on my guidance... steering wheel and pedals is not easy, especially from a distance ..."*, and RO7 said *"... Absolutely yes! ... The human-machine mix to deal with issues can fit because today's vehicles are very sensing ... I believe such an approach can be very helpful for smooth teleoperation ..."*

Four of the above users (P6, P9, P12, P14) added that using commands is context-dependent. P6 shared that *"... It depends ... On the one hand, remote driving is inefficient and has latency issues and safety concerns, so you can monitor more vehicles if you have just the commands. On the other hand, not all scenarios can be handled by discrete commands. We can use discrete commands in 7 out of 10 cases; in the other 3, we will still need manual driving ..."*. Finally, one participant (RO13) said that, in her opinion, remotely controlling an AV using discrete commands can be possible only in an environment where all the vehicles are autonomous.

*Relevance of the Suggested Commands*

The RO must be willing and capable of using each suggested command to resolve a scenario. For instance, four participants (RO4, RO11, RO12, RO14) were reluctant to use the U-turn command to resolve the "Busy Junction" use case. Participants motivated their choice by claiming that performing a U-turn does not allow them to integrate into the congested traffic and continue their planned route. Furthermore, not every situation with congested traffic will enable a "U-turn" or "Reroute." For instance, RO5 noted that *"... A vehicle that should do 'Reroute' in such situations [entering busy intersection] isn't a good system design ... Additionally, it's not practical because you might enter heavy traffic anywhere ..."*. Thus, every high-level command that is suggested to the RO by the assisting AI agent should be contextually related to the presented scene and enable its resolution. Other commands can be located in a less prominent section of the UI.

*Injection of Control vs. Immediate Action*

According to RO4, it is essential to differentiate between two major types of commands. The first type is "Injection of Control," where RO receives a high-level decision and delegates its execution to the AV to be executed autonomously. Such a command can be "Bypass from Left" or "Plot Alternative Route" (see [65]). The second type is a command that executes immediately. For instance, when entering a busy junction (see [64]), the RO might repeatedly use the "Progress Slowly" command. Several users (RO1, RO2, RO6, RO7, RO11) suggested to refine the interaction with this command. For instance, RO6 shared his confusion: *"... When we say 'progress slowly,' is it going to slide and stop, or is it going to move a fixed distance and stop? ..."*. It is also important to note that *"... [we] are not trying to drive in real-time with buttons ..."*. RO4 raised the same issue when discussing the "Plot Alternative Route" command. She noted that the path plotting routine requires the RO to *"... create something static in a dynamic world [which is a problem] ..."*. Therefore, creating a hierarchical structure of commands is imperative when designing a command-based interface. In each case, the control scheme should be well thought through and the interaction carefully crafted.

*Camera and Map Control as a Tool to Increase Situational Awareness*

Participants requested to gain increased control over the camera perspectives and the map. Several users (RO1, RO2, RO3, RO7, RO10, RO14) mentioned the need to enable a 360-degree view, specifically the rear camera, in some cases (e.g., "U-Turn" when a tree blocks the road). Additionally, several users (RO6, RO7, RO8, RO13, RO14) desired the ability to explore the vehicle's environment using cameras that they could control. P7 shared with us that *"...One of the most* *important things is to control what I see completely and also the convenience of this aspect ... I need specific angles such as 'In Vehicle,' 'On the side,' 'Rear View,' etc. in the fastest possible way. I [also] need all a PTZ camera at hand ...".* RO2 also emphasized the need for an in-vehicle camera, which will enable video communication with the passengers when needed. Additionally, participants (RO1, RO6, RO8, RO12, RO13) wanted to have the option to see and explore the map. For example, P13 said that *"... [she] would like to have an option to move the map left and right [pan] ...".*

### Providing Context to Increase RO's Feeling of Control

Several methods to increase RO's feeling of control were raised during the study. RO1 repeatedly stated that *"... the map view gives [her] a feeling of control ..."* and RO6 wanted to see a small map window near the video feed at all times. Additionally, RO6 and RO11 expected to see a larger picture of the fleet, and RO11 shared that she wants to see *"... a bunch of vehicles assigned to [her since she] was used to first seeing the [whole] fleet screen, and only then select a vehicle and connect to [it]...".* Moreover, RO10 explicitly stated that she feels less control when she operates the AV using discrete commands. Another mentioned issue was related to the "Reroute" command. Participants wanted to know how the command influences the estimated arrival time and the route to the final destination. Finally, RO1 and RO3 wished to see the rear camera while using the "U-Turn" command. Therefore, designers should strive to increase the RO's feeling of control in all possible ways. For instance, exposing the RO to the operational context, such as showing a queue of all RO's tasks and a map of a larger area.

### Dominance of the Video Feed

Several participants shared that the real-time video feed is very dominant and attracts most of the RO's attention resources. An observation that is bolstered by existing research on this topic [63] states that when driving remotely, the RO is in a constant state of high alert. This fact has at least two major design implications. First, several participants (RO2, RO5, RO10) shared that the video scene is so dominant that peripheral UI elements (e.g., a warning sign) may be missed. Therefore, it should be considered that the peripheral UI elements be reduced to the necessary minimum and placed on top of the video feed when appropriate. Additionally, placing more information on top of the video feed, where the RO's gaze is focused, might be beneficial. RO6 suggested that the augmented reality layers reflect more feedback regarding the high-level commands. For instance, if the RO selected the "Progress Slowly" command, the interface should show what will be the progression distance.

### The Danger of Gamification

In our design, we chose a 3$^{rd}$ person-view perspective for the real-time video feed, similar to various driving games. Additionally, we were guided by the Material Design[7] guidelines. Such design choices made several participants (RO4, RO6, RO14) raise concerns that ROs might treat teleoperation as a game rather than dangerous maneuvering that requires responsibility and caution. Therefore, we believe that the look and feel of future teleoperation interfaces should be designed as realistically as possible to create a feeling of immersion in the remote environment, which might lead to more caution and responsibility.

### Tele-driving vs. Tele-assistance

When discussing the "Busy Intersection" scenario, several participants (RO4, RO5, RO12, RO14) preferred to use a steering wheel and pedals rather than merging into traffic using high-level commands. RO4 said: *"... In this situation, I*

---

[7] https://m2.material.io/design/guidelines-overview

*don't want to resolve the situation using semi-autonomous commands ... I want to get to a state where I do an action, and the vehicle immediately executes it, and I also want to be able to change the action. This might not be only a stop because maybe a motorcycle enters the junction, and I want to do an evading maneuver ..."*. The above scenario highlights the fact that in some cases, tele-driving, despite its challenges, might be advantageous over tele-assistance and vice versa. Therefore, not only should tele-assistance user interfaces need to enable easy and straightforward transitioning to tele-driving, but the system should also be able to suggest which operation method is preferable for which edge case.

*5.2.2 Quantitative evaluation*

We used the PSSUQ questionnaire to measure overall usability. The aggregated PSSUQ questionnaire results are presented in Tables 3 and 4. Since our study was based on a click-through prototype but not a fully dynamic implementation, we removed questions 7 and 9 from the questionnaire because they were irrelevant. The overall average usability score (OVERALL) of the prototype was 2.775 (1 - high usability, 7 - low). Additionally, we calculated the associated subscales of usefulness, information quality, and interface quality [40]. The results are presented in Table 3 and compared to benchmark scores provided by Sauro and Lewis, who aggregated data from 21 studies using the PSSUQ questionnaire [59]. The results show overall good scores with a high correlation to the benchmark.

Table 3: Descriptive and inferential statistics regarding usability.

| Construct | M | Benchmark Values | SD |
|---|---|---|---|
| Overall score (OVERALL) | 2.775 | 2.82 | 1.175 |
| System usefulness (SYSUSE) | 2.357 | 2.8 | 1.046 |
| Information quality (INFOQUAL) | 3.464 | 3.02 | 1.345 |
| Interface quality (INTERQUAL) | 2.714 | 2.49 | 1.264 |

Table 4: PSSUQ questionnaire results.

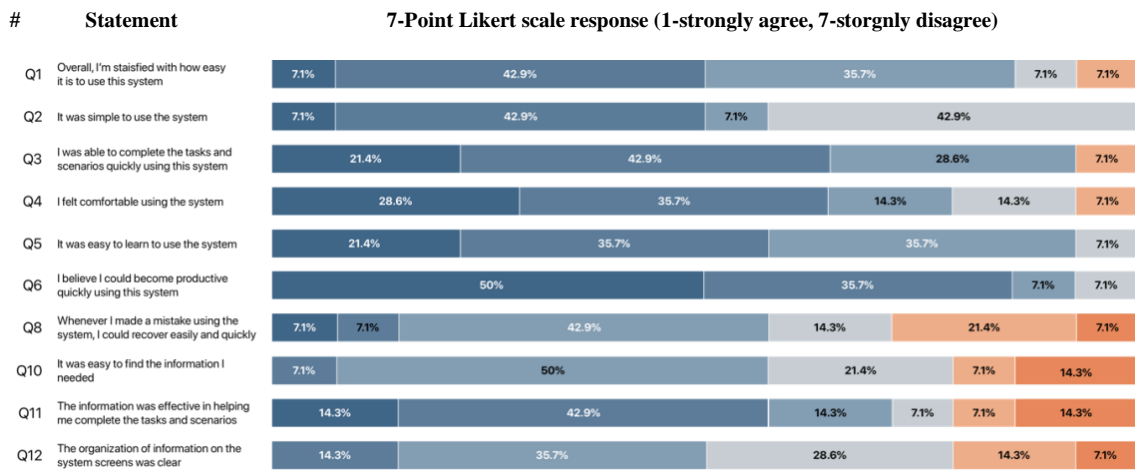

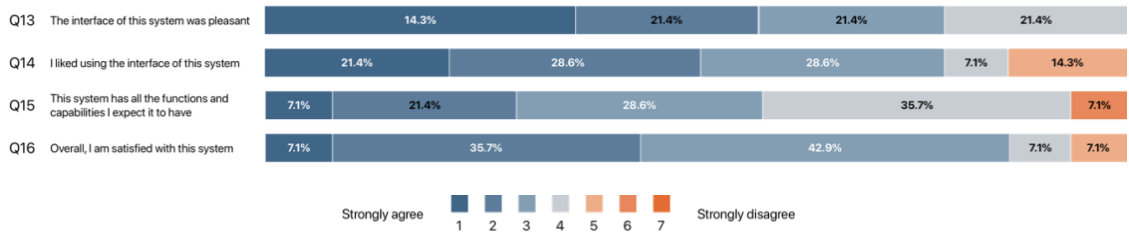

The Van Der Laan (VDL) questionnaire [38] is a tool to measure systems acceptance on two dimensions: a *Usefulness* dimension and an affective *Satisfaction* dimension. Practical aspects of the system are reflected in the *Usefulness* score, and the pleasantness is mirrored in the *Satisfying* score. Questionnaire scores are calculated on a scale from -2 (low) to 2 (high). We calculated both of the above constructs and received the following results: *Usefulness* = 0.885 (2 - useful, -2 - useless) and *Satisfying* = 1.053 (2 - satisfying, -2 - unsatisfying). The average standard deviation across categories was SD=0.859. These results indicate that the system was relatively well accepted by the participants.

## 6 DISCUSSION

Here, we discuss the high-level issues that stem from our study that are relevant to the larger research community, the industrial complex, and the regulators, as well as issues that should be explored in future research.

*Control Ownership*

In tele-driving, the transition of control between the AV and the RO is well-defined: after a take-over request [3,5,46], the RO assumes control and continuously drives the vehicle (via a steering wheel and pedals or other methods). Once the edge-case scenario is resolved, the control is returned from the RO to the AV. However, in tele-assistance, the situation is more complex because the RO provides high-level instructions while the AV is still responsible for its low-level maneuvering. In some cases, it is more appropriate that the RO overrides AV decisions; in others, the opposite is true. For instance, if an AV is stuck because a continuous separation lane does not enable it to overtake an animal that blocks its main course, it is appropriate to cross the separation lane, given the RO's permission.

On the other hand, if the RO guides the AV to continue its movement into a wall, the AV should not allow it. We believe that when there is a logical constraint (e.g., a shadow [64]), the RO should be able to override AV's safety envelope. On the other hand, when there is a clear physical obstacle (e.g., a deep puddle [64]), the RO should not be able to override the AV's operational design domain (ODD)[8].

Control ownership is related to the legal aspects of remote operation of AVs. For instance, crossing a continuous separation lane is an illegal action, and it is important to determine who is liable in the event of an accident. Additionally, ensuring accountability for decisions made by the autonomous system or its remote operator, especially in life-threatening situations, raises ethical concerns. Therefore, collaboration between various stakeholders (vehicle manufacturers, national authorities for road safety, lawmakers, etc.) is essential for the development of a comprehensive regulatory framework.

---

[8] SAE J3016 defines ODD as "Operating conditions under which a given driving automation system or feature thereof is specifically designed to function, including, but not limited to, environmental, geographical, and time-of-day restrictions, and/or the requisite presence or absence of certain traffic or roadway characteristics."

*Implementational Feasibility of a Sensor-Fusion-Based Scene*

All participants reported that the suggested ("World View") camera perspective (see Figure 3) is helpful because it provides a higher feeling of control (RO1) and increases remote situational awareness (RO2). Additionally, RO7 suggested improving this view by adding data from other sources (e.g., other vehicles) and providing system notification (e.g., a car approaching the junction at high speed). Saying the above, RO14 noted that the presented *"... perspective is an amazing view, but you don't have it in reality. Sensor fusion will not give the presented 2D or 3D views ..."* because LIDARs are limited in their distance and angels and various objects block their laser beams. At the same time, P14 mentioned that after implementing something similar for his business unit, *"... the users stopped using the cameras and used only this view ..."*. Based on the discussion above, it is evident that a "World View" perspective can be very useful to ROs, however, implementational complexities are yet to be resolved.

*Does Tele-Assistance Save Time and Increase Safety?*

One of the assumptions of this research work is that tele-assistance might shorten the duration of a teleoperation session, thus making teleoperation services more affordable and bringing closer a full-scale deployment of AVs on public roads. However, after issuing a high-level command, the RO must also monitor its execution till the scenario is resolved. Therefore, despite the promising approach, it is unclear whether tele-assistance saves time, how much, and in which use cases.

One open question is exploring whether tele-assistance can enable the one-to-many paradigm. While in tele-driving, the RO always has to have her hands on the steering wheel. This is not the case in tele-assistance because the AV is responsible for all the low-level maneuvers and their safety. Therefore, after issuing high-level guidance and waiting for its execution, the RO could resolve the following issue during the monitoring stage of the first use case.

Another pertinent question is whether tele-assistance indeed increases safety. What happens if the RO gives a command that overrides AV's ODD (e.g., perform an obstacle bypassing maneuver despite a continuous separation lane) and the AV does an accident? Should a vehicle progress to a destination that was marked on the sidewalk? While all the above questions should be carefully thought through in future design iterations, we believe that when there is a logical deadlock, the RO should be able to override the ODD (see example above). At the same time, if there is a physical obstacle (e.g., the RO guides an AV to drive into a wall), the AV should use perception and ensure its safety. Answering the above questions is left for future work explorations.

*Are High-Level Commands Flexible Enough?*

This research focused on resolving edge-case scenarios using a set of contextual high-level commands. As mentioned above, such an approach has many advantages. However, our research also unveiled the flexibility of "Point and Go" and "Plot Alternative Route" commands, in which the RO can point anywhere on a map and ask the vehicle to follow. Thus, one of the valid questions is whether plotting points on the map is sufficient. In some scenarios, issuing a high-level command, such as "Turn Right," is much more efficient than plotting a route. At the same time, in other cases, e.g., bypassing an ambulance in a traffic jam, using a more flexible interaction will be better. Therefore, we believe the best approach is to have both methods. In future research, we intend to ask users to use the designed tele-assistance UI to resolve several edge-case scenarios and, among other things, infer which commands were used for each use case.

## 6.1 Limitations

This study had several limitations. First, we used an InVision-based prototype based on simulation-based images but not a fully developed system. This fact sometimes confused users, who expected a more responsive system that reacts to any of their actions. For example, it was hard to fully understand the control ownership scheme because the feedback from a live simulation was missing. Because of the same reason, users could not estimate other vehicles' speeds. Another example was users' inability to plot a route freely since they had to press only on predefined screen areas. Finally, because of the prototype's static nature, it was impossible to cover all interaction flows. Thus, error handling, for example, was mostly missing from the prototype.

A second limitation was users' bias towards previously used systems. In some instances of the study, it was evident that users compared the evaluated interface with previously used systems. Their previous experience was often advantageous, but users sometimes leaned toward previously known paradigms. For instance, RO11 suggested having a queue of AVs waiting for teleoperation because that is what she used in the past, even though such a design might distract the RO from the task at hand, and the fact that this feature might be more appropriate for a managerial role within the teleoperation center. Similarly, RO2, who had experience in drone operation, suggested having the "Indian Circle" feature in which a camera rotates around a particular spot in space. While such a feature could significantly increase RO's situation awareness, it is appropriate only for flying vehicles.

Finally, in real-world systems, teleoperators will probably have rigorous training using the systems they will use to perform their tasks. In contrast, in our study, there was no proper system learning. Thus, some of the participants' comments might not require design changes but can be resolved by training.

## 6.2 Future Work

In the next phase of this research work, we plan to implement the presented solution and perform a comparative study between tele-driving and tele-assistance paradigms while simulating the assisting AI agent using a Wizard-of-Oz technique. More specifically, we intend to simulate several edge-case scenarios and ask nonexpert users to resolve each scenario using both methodologies. This will enable us to validate whether tele-assistance reduces the teleoperation time, improves cognitive load and situation awareness, and enhances safety.

Another promising research direction is the involvement of the passenger in remote scenario resolution. In addition to the human intelligence of the RO and the artificial intelligence of the AV, the passenger also possesses human intelligence, which can be helpful in some scenarios [12]. For instance, the passenger can add details about the situation that are not visible to the RO, or she can get out from the AV and move a tree branch that blocks the road. Furthermore, providing the passenger with a user interface to resolve problems has the potential to avoid latency issues. However, not all passengers can make safe and responsible decisions (e.g., kids, sick people, drunk people, etc.). Therefore, while involving the passengers in scenario resolution might be advantageous, such an approach should be carefully examined.

## 7 CONCLUSION

Our research explores the tele-assistance paradigm, a transformative approach to remote operation of AVs that seeks to address the limitations of conventional teleoperation methods. By developing and rigorously testing a command-based user interface prototype, we highlighted the potential of high-level commands to streamline AV operation, particularly in scenarios where manual driving may prove inefficient or unsafe. Our study reveals a strong preference among expert teleoperators for discrete, high-level commands but also emphasizes the need for adaptable control modes to handle diverse situational demands effectively. The research identifies critical areas for refinement, such as integrating enhanced situational awareness through advanced camera and map controls, essential for improving operator confidence and feeling of control. Furthermore, our findings advocate for intelligent interface designs that dynamically suggest the most appropriate operational mode, optimizing safety and efficiency. This research establishes a robust foundation for future developments, guiding the evolution of interfaces that will enhance the interaction between human operators and autonomous systems. As AV technology progresses, our study provides an important stepping stone toward more sophisticated and responsive tele-assistance systems.

68. Tao Zhang. 2020. Toward Automated Vehicle Teleoperation: Vision, Opportunities, and Challenges. *IEEE Internet of Things Journal* 7, 12: 11347–11354. https://doi.org/10.1109/JIOT.2020.3028766

69. John Zimmerman and Jodi Forlizzi. 2011. Research through Design: Method for Interaction Design Research in HCI. *Chi 2011*: 167–189.

**APPENDIX A**

The following table summarizes various usability issues that were discovered during our study and are specific to the presented user interface.

Table 5: Major unveiled usability issues.

| Usability Category | Issue Description | Participants |
|---|---|---|
| Augmented Reality Layers | Most (11) participants did not understand the design and the purpose of the square, red, semi-transparent layer in front of the AV, which was designed to show that the AV has stopped (Figure 4, area 2). Participants suggested various design approaches to resolve the misunderstanding: (1) Using a different color, (2) Convert the color of the projected green path to red, (3) Creating a red halo around the AV, (4) Add a large red cross in front of the AV, (5) Lighting the breaks lights, (6) Adding a red border to the entire simulation screen [when it turns green the AV can continue its movement], and (7) Making the rectangle wider. When it turns green the AV can continue its movement. | RO1, RO2, RO4, RO5, RO7, RO8, RO10, RO11, RO12, RO13, RO14 |
| Unclear UI Elements | Some of the icons were not entirely clear. A major confusion was caused by the "Microphone" and the "Honk" icons (see Table 1). Eleven participants thought that a "Microphone" is a "Mute" and five participants expected the "Honk" icon to activate an external microphone. Additionally, the "Map" icon was thought to be a "Full Screen" icon by six participants. | RO1, RO2, RO3, RO4, RO5, RO6, RO7, RO8, RO10, RO11, RO12, RO13, RO14 |
| Location, size, and color of UI elements on the screen | ▪ The "stop" button was not clear to several users. Users suggested to move the "stop" button to the right panel, make it much bigger, change the "hand" icon to a "STOP" text and enable toggling between "STOP" and "GO". Additionally, RO7 noted that there should also be a big physical "emergency stop" button, which will be separated from the UI and enable visceral stoppage. | RO4, RO5, RO6, RO9, RO11, RO12 |
| | ▪ Several users emphasized the need to co-located related UI components. Specifically, four users suggested to move the yellow guidance pop-up to the right panel because the guidance is related to the commands. | RO3, RO4, RO6, RO8, RO9, RO11 |
| Visibility of Components | ▪ The pop-ups at the upper part of the screens were not visible enough to all users. RO2 suggested instead to show a prominent pop-up in the center of the screen, which can be collapsed to its current location after the RO see it. RO11 suggested to make the pop-ups bolder and bigger. | RO2, RO3, RO4, RO8, RO9, RO10, RO11, RO12 |
| | ▪ Several users said that the "Control Owner" status should be more visible and some suggested to rename them. For instance, RO9 suggested "Auto" instead of "Vehicle" and "Manual" instead of "Remote Driving". | RO5, RO9, RO11, RO13 |
| Interaction and UX | ▪ The interaction with the "Progress Slowly" command should be refined. This command is different than other commands since using this command the RO provides a more granular control of the AV. Therefore, several suggestions were made: (1) "Click & Hold", (2) "Slide [until stopped]", and (3) "Progress a predefined distance". | RO1, RO2, RO6, RO7, RO11 |
| | ▪ The design of the "Gap Keep" command was not clear to several users. Specifically, canceling the command was a challenge and users also lacked various contextual adjustments (e.g., gap keeping distance). Additionally, several logical issues were raised. For instance, RO3 noted that there might be a problem if the AV needs to stop on a "stop" sign and also keep the gap from the vehicle ahead. RO8 noticed that it might be a problem to do the gap keeping in real- | P1, P2, P3, P5, P6, P7, P8, P13, P14 |
| | time since the vehicle ahead might move drive away while the RO is busy making the selection. | |
| | ▪ In some cases participants expected to have less clicks to achieve a certain result. For instance, several users noted that canceling the "Gap Keep" command was cumbersome. Another example is too many confirmation buttons. | RO2, RO5, RO9, RO10, RO13 |
| Newly Suggested Elements | ▪ Access to low-level controls of the AV: (1) Hazard lights, (2) Winkers, (3) AC controls, (4) Oil level, etc. | RO2, RO6, RO7, RO14 |
| | ▪ Access to more details about the AV and the passengers: (1) Vehicle ID, (2) Door status, (3) Number of passengers, (4) How many seatbelts are buckled, (5) Passenger's name, (6) Vehicle type, etc. | RO5, RO6, RO7, RO8, RO12 |
| | ▪ Making visible the AV's telemetry: (1) Speed, (2) Acceleration, (3) RPM, (4) Visualization of gas and breaks pedals, etc. | RO2, RO6, RO7, RO13 |
| | ▪ Adding an option to measure width of objects (e.g., an obstacle) and distances between various objects in the remote scene (e.g., passage width between buildings). | RO3, RO10 |
| | ▪ Event logs (e.g., footage before and after the event) | RO2, RO6 |
| | ▪ Physical emergency stop button ("mushroom") | RO5, RO7 |
| | ▪ Adding route-related information when rerouting: new time and distance to destination. | RO1, RO2 |
| | ▪ Showing a clock in the UI to understand how long the passenger is waiting. | RO6 |
| Utilization of Hardware Affordances | Enabling various touch interaction techniques. For example, "pinch" for zooming and "panning" to move the scene around. RO12 suggested to use the mouse wheel to zoom in and out. | RO6, RO12 |
| Traffic Signs | Overall, the traffic signs were clear to the majority of the participants, but RO6 noted that he isn't familiar with some of them. Therefore, traffic signs should be localized according to the country, where the AVs are driving. | RO6 |